\documentclass[preprint,amsmath,amssymb,pre,superscriptaddress]{revtex4-1}
\usepackage{color,longtable}
\usepackage{graphicx}
\usepackage{amssymb,amsfonts,amsmath}
\usepackage{here}

\begin{document}
\title{Cooperative Ion Conduction Enabled by Site Percolation in Random Substitutional Crystals}
\author{Rikuya Ishikawa}
\affiliation{%
Department of Physics, Tokyo Metropolitan University, 1-1 Minamioosawa, Hachiouji-shi, Tokyo 192-0397, Japan
}
\author{Kyohei Takae}
\affiliation{%
Department of Mechanical and Physical Engineering, Tottori University, 4-101 Koyama-cho Minami, Tottori 680-8550, Japan
}%
\author{Rei Kurita}
\affiliation{%
Department of Physics, Tokyo Metropolitan University, 1-1 Minamioosawa, Hachiouji-shi, Tokyo 192-0397, Japan
}%
\date{\today}
\begin{abstract}
Efficient and safe energy storage technologies are essential for realizing a sustainable and electrified society. Among the key challenges, the design of superionic conductors for all-solid-state batteries often faces a fundamental trade-off between stability and ionic conductivity.
Random substitutional crystals, where atomic species are randomly distributed throughout a crystal lattice, present a promising route to overcome this trade-off.
Although the importance of cooperative motion in ion conduction has been pointed out, there is a lack of understanding of the relationship between mesoscale structural organization and macroscopic conductivity, limiting the rational design of optimal compositions.
Here, we systematically investigate the ionic conductivity of rock salt random substitutional ionic crystals Li$_x$Pb$_{1-2x}$Bi$_x$Te as a function of Li concentration $x$ using molecular dynamics simulations.
We find that ionic conductivity increases sharply once the $x$ exceeds a critical threshold, without disrupting the underlying crystal structure.
Strikingly, this threshold aligns with the site-percolation threshold predicted by percolation theory.
Our findings establish ion percolation as a universal design principle that reconciles the trade-off between conductivity and stability, offering a simple and broadly applicable strategy for the development of robust, high-performance solid electrolytes.
\end{abstract}


\maketitle

\section{Introduction}
Designing materials that enable fast and efficient ionic transport is a central challenge in physics, chemistry, and materials science, with far-reaching implications for energy storage, catalysis, and sensing technologies. Among these, all-solid-state batteries have emerged as a promising platform for next-generation energy storage due to their intrinsic safety and potential for high energy density.
 A key requirement for advancing such technologies is the discovery of superionic conductors and a deeper understanding of their ionic conduction mechanisms~\cite{Nitta2015, Takada2013, Xiao2020}. Various lithium-containing oxides and sulfides have been reported to exhibit high ionic conductivity~\cite{Kamaya2011, Li2023}, making them promising candidates for next-generation battery applications.

For all-solid-state batteries to be viable in practical settings, high ionic conductivity must be coupled with excellent chemical stability. Stability challenges include maintaining strong adhesion at the electrode-electrolyte interface without undesirable chemical reactions~\cite{Xiao2020, Richards2016}, and preventing the release of toxic gases upon air exposure~\cite{Muramatsu2011}.
One widely explored approach is doping, where ions of different valence states are introduced to increase the concentration of vacancies and mobile ions~\cite{Murugan2007, Michael2008, Miyazaki2018, Berardan2016, Yu2020}. 
However, excessive doping often compromises structural stability and, in some cases, can even reduce ionic conductivity due to overcrowding of ions within the lattice~\cite{Ramakumar2015, Inaguma1996, Pu2025}. 
Despite numerous proposed strategies, the simultaneous realization of high conductivity and stability remains elusive, complicating the rational design of high-performance materials.

To overcome this trade-off, we focus on a class of materials we refer to as random substitutional ionic crystals (RSICs), in which ions are randomly substituted throughout the lattice. 
RSICs resemble high-entropy alloys~\cite{Senkov2017, Yeh2004, Cantor2004} and entropy-stabilized compounds~\cite{Rost2015}, which are known for their exceptional thermodynamic stability due to large configurational entropy. These materials also benefit from element-specific synergistic effects that enhance properties such as mechanical strength~\cite{Okamoto2016}, catalytic activity~\cite{Kusada2020, Kusada2022}, thermoelectric performance~\cite{Jiang2022, Yamashita2021}, and superconductivity~\cite{Mizuguchi2019, Kasem2020, Kasem2022, Matsumoto2024}.
While materials with strong ionic bonding, such as conventional ionic crystals, provide excellent structural stability and high ion density, they inherently exhibit low ionic conductivity~\cite{Aniya1992, Katsumata2002}. However, the unique structural characteristics of RSICs may enable both high stability and high ionic conductivity, potentially overcoming the conventional trade-off.

To explore the potential of RSICs in ionic transport, we previously performed molecular dynamics (MD) simulations of the rock salt high-entropy material AgInSnPbBiTe\textsubscript{5}~\cite{Mizuguchi2023, Ishikawa2024}, we found that In\textsuperscript{+} spontaneously forms Frenkel defects, promoting atomic diffusion~\cite{Ishikawa2024}. This observation suggests that defect formation and carrier mobility may be tuned through careful selection of ionic species and control over their spatial distribution.
Furthermore, in recent years, it has been suggested that the cooperative motion of ions is important in ionic conduction~\cite{Esaki2025, Burbano2016, Morgan2014, Morgan2021, He2017}.
However, the connection between mesoscale carrier distribution, such as their spatial arrangement and connectivity, and macroscopic ionic transport remains largely unexplored. Bridging this gap is essential for developing design strategies that balance conductivity and stability.

Here, we investigate the ionic conductivity of rock salt RSICs containing carrier ions with various charges, focusing on the relationship between carrier concentration and bulk conductivity. 
In this study, we mainly investigated the rock salt Li$_x$Pb$_{1-2x}$Bi$_x$Te system, which is derived from the rock salt AgInSnPbBiTe$_5$ compound. In this model, the monovalent cation is replaced by Li$^+$. Because Li$^+$ has a small ionic radius, it is expected to exhibit enhanced ionic conduction.
Our MD simulations reveal a sharp increase in conductivity once the carrier concentration exceeds a critical threshold, without disrupting the underlying crystalline structure. Remarkably, this threshold aligns with the site-percolation threshold predicted by percolation theory~\cite{Stauffer1992}, directly linking microscopic carrier organization with macroscopic transport behavior.
This percolation-based framework could be broadly applicable across a range of multicomponent ionic systems, providing new insights into the design of robust and high-performance solid-state electrolytes.

\section{Methods}
We performed molecular dynamics (MD) simulations on a simple cationic three-component system, Li$_x$Pb$_{1-2x}$Bi$_x$Te, as a model for RSIC. To ensure overall charge neutrality, the number of Li$^+$ and Bi$^{3+}$ in the system must be equal, as required by charge conservation.
MD simulations were performed to investigate the mechanisms of ionic conduction under external electric fields. 
The simulation procedure largely follows the method described in Ref.~\cite{Ishikawa2024}. 
Particles $i$ and $j$ were subjected to interaction potentials including the WCA potential $U_{\mathrm{WCA}}$ and the Coulomb potentials $U_q$. 
\begin{eqnarray}\label{EoM}
m_i \frac{d^2 \vec r_i}{dt^2} &=&q_i\vec E+ \sum_{i\neq j} -\vec \nabla (U_{\mathrm{WCA}} (r_{ij}) + U_q (r_{ij})) \\ 
U_q (r_{ij}) &=& -\frac{k q_i q_j}{r_{ij}}
\end{eqnarray} 
\begin{eqnarray}
U_{\mathrm{WCA}} (r_{ij}) &=& \left\{
\begin{array}{ll}
4\epsilon \left[ \left(\frac{d_{ij}}{r_{ij}} \right)^{12} - \left(\frac{d_{ij}}{r_{ij}} \right)^6 \right] +\epsilon & (r_{ij} < 2^{1/6}d_{ij}) \\
0 & (r_{ij} \ge 2^{1/6}d_{ij})
\end{array}
\right.
\end{eqnarray}
where $\vec r_i$, $r_{ij}$, $m_i$, $\epsilon$, $k$, $q_i$, $\vec E$ are the coordinate of particle $i$, the center-to-center distance from particle $i$ to $j$, the mass of particle $i$, the energy coefficient, the Coulomb constant, the charge of particle $i$, and external electric field, respectively. 
$d_{ij} = (d_i + d_j)/2$, where $d_i$ is a diameter of particle $i$. 
The length and mass of the particles were normalized by the ionic diameter and mass of Te$^{-2}$ ($d_{\mathrm{Te}} = 4.42~ \mathrm{\AA}$~\cite{Shannon1976}, $m_{\mathrm{Te}} = 127.6 ~\mathrm{g/mol}$~\cite{Thomas2022}), respectively. 
Therefore, ionic diameter of Li$^+$, Pb$^{2+}$, Bi$^{3+}$, and Te$^{2-}$ ions were 0.344, 0.538, 0.466, and 1.00, respectively~\cite{Shannon1976}. Mass of Li$^+$, Pb$^{2+}$, Bi$^{3+}$, and Te$^{2-}$ were 0.0543, 1.61, 1.63, and 1.00, respectively~\cite{Thomas2022}. 
Here, the interatomic interaction $\epsilon = 295 k_B ~\mathrm{J}$ are assumed to be constant. 
Subsequently, time, temperature, pressure, and electric field were measured in units of $t_0 = \sqrt{m_{\mathrm{Te}}d^2_{\mathrm{Te}}/\epsilon} = 3.18~\mathrm{ps}$, $T_0 = \epsilon/k_{\mathrm{B}} = 295~\mathrm{K}$, $P_0 = \epsilon/d_{\mathrm{Te}}^3 = 4.71\times 10^7~\mathrm{Pa}$, and $E_0 = \sqrt{\epsilon/d_{\mathrm{Te}}^3} = 6.61 \times 10^{8}~\mathrm{V/m}$, respectively. 
In this study, the temperature was set to $T = T_0$. We primarily set the system pressure to $P = P_0$.
We also performed simulations at a lower pressure of $P = 0.01P_0$, which corresponds approximately to atmospheric pressure, and the results were found to be qualitatively unchanged. 
We integrated Eq.~\ref{EoM} using the leap-frog method with the time step width being 0.002. 
The unit charge $q_0$ and ionic conductivity $\sigma_0$ was normalized by $q_0 = \sqrt{\epsilon d_{\mathrm{Te}}} = 1.40\times10^{-20}~\mathrm{C}$ and $\sigma_0 = \frac{q_0^2t_0}{m_{\mathrm{Te}}d_{\mathrm{Te}}^3} = 0.34~\mathrm{S/cm}$. For this study, we set $k$ = 1 and $e = 5q_0$, where $e$ is the elementary charge. 
In our previous work on the rock salt AgInSnPbBiTe$_5$ system, simulations using the same potential showed qualitative agreement in properties such as the longitudinal sound velocity and melting temperature with experiments~\cite{Ishikawa2024}. This suggests that simulations of rock
salt telluride systems can be reasonably performed with this potential.
The total number of particles was set to $N = 24^3 = 13824$, arranged in a rock salt crystal structure, and the cation types were randomly arranged. 
Initial conditions were set to run for a time interval of 200 in the NPT ensemble, with $P$ and $T$ controlled by the Andersen barostat and the Nose-Hoover thermostat, respectively~\cite{Frenkel2002}. 
The stress and $T$ were then held constant using Parrinello-Rahman's method~\cite{Parrinello1981} and Nose-Hoover thermostat, and an external electric field was applied. After applying the electric field, we analyzed the properties when a steady state was reached after waiting a time interval of 300.
We chose the thermostat and barostat time constant as 0.1 and 1, respectively.
Coulomb force was computed by Ewald summation method~\cite{Ewald1921, Frenkel2002}. 
By choosing the Ewald parameter $\alpha =0.6$, the cutoff length in the real space 6 $d_{\mathrm{Te}}$ and the cutoff wavenumber 11(2$\pi$/$L$) where $L$ is the system length, the root mean square error in the force became less than $10^{-4}$. 
We adopt periodic boundary conditions. 
All simulations were performed using a self-developed molecular dynamics code.

Ionic conductivity was calculated using the following equation
\begin{eqnarray} \label{sigmaE}
\sigma &=& \frac{\langle J \rangle_t}{E} \\
\langle J \rangle_t &=& \left\langle \sum_i \frac{q_i v_i}{V} \right\rangle_t
\end{eqnarray}
where $J$, $v_i$, and $V$ are current density, velocity of particle $i$, and volume, respectively. 
$\langle \rangle_t$ represents time average. 
To reduce the effect of fluctuations, $v_i$ was defined as $v_i=\frac{r_i(t+\Delta t)-r_i(t)}{\Delta t}$ and $\Delta t=10$ as the time scale on which particles begin to move. 
The directions of the $E$ are (001), (101), and (111) with respect to the crystal plane, and $J$ and $v_i$ are the same directional components as the electric field direction.
Thus, $\sigma$ corresponds to the ionic conductivity in the $E$ direction.
Note that the ionic conductivity in the direction perpendicular to the electric field direction was found to be almost zero.

\section{Results}
\begin{figure}[htbp]
\centering
\includegraphics[width=8cm]{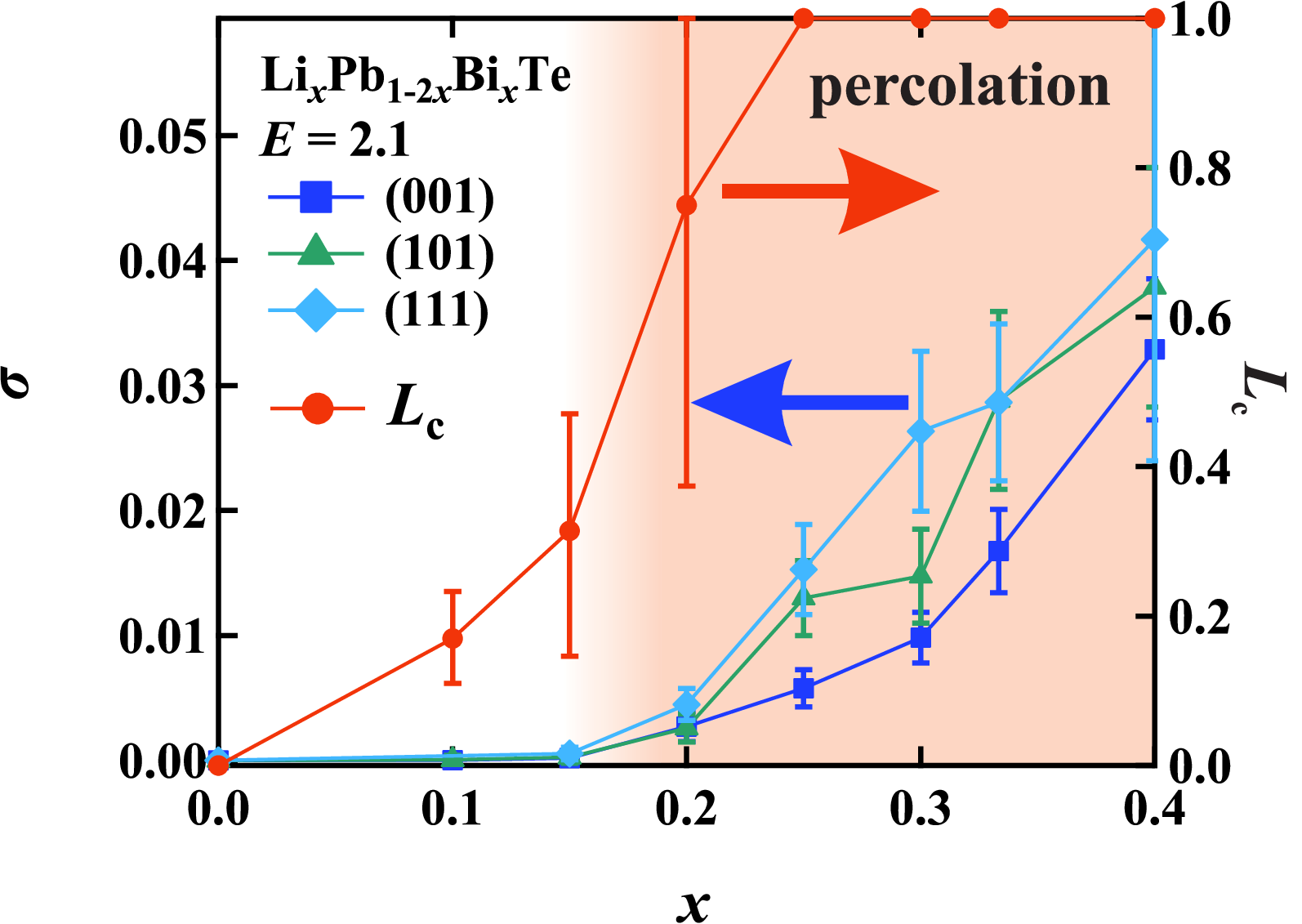}
\caption{The $x$-dependence of ionic conductivity $\sigma$ (left axis). Square, triangle, and diamond symbols represent $\sigma$ under electric fields applied along the (001), (101), and (111) directions, respectively. Symbols and error bars indicate time-averaged values and their standard deviations. Independent of the electric field direction, $\sigma$ begins to increase sharply from $x = 0.2$. This increase can be attributed to changes in the internal Li$^+$ configuration occurring at $x = 0.2$.
The $x$-dependence of the length $L_c$ of the largest Li$^+$ cluster is also shown (right axis). $L_c$ is normalized by the maximum simulation box length, with $L_c = 1$ indicating percolation. Circle symbol corresponds to time-averaged values. Error bars correspond to maximum and minimum values of $L_c$. 
$L_c$ begins to increase from $x = 0.2$, indicating the onset of percolation, and $L_c$ = 1 beyond $x = 0.25$. 
}
\label{sigma}
\end{figure}

We performed molecular dynamics (MD) simulations on a simple cationic three-component system, Li$_x$Pb$_{1-2x}$Bi$_x$Te, as a model for RSIC. In these simulations, we considered only excluded volume effects and Coulomb interactions, neglecting other interactions to focus on the primary factors governing ion transport.
We investigated the dependence of ionic conductivity $\sigma$ on the Li$^+$ ion concentration $x$ under applied external electric fields $E$. 
As shown in the Supplementary Information [\onlinecite{supplement}], the radial distribution functions for $E = 0$ and $E = 2.1$ are identical, indicating that the crystal structure is stable under the applied electric field.
Figure~\ref{sigma} shows the variation of $\sigma$ with $x$ at an applied field strength of $E = 2.1$. 
Even when $E$ is varied, the $x$-dependence trend of $\sigma$ remains unchanged; only the magnitude of $\sigma$ varies. Details of the $E$-dependence of the magnitude of $\sigma$ are provided in the Supplementary Information [\onlinecite{supplement}].
Squares, triangles, and diamonds represent measurements under electric fields directed along the (001), (101), and (111) crystallographic axes, respectively.
We find that $\sigma$ remains quite low at small $x$ but begins to increase sharply once $x$ exceeds approximately 0.2. 
Note that we also estimated the ionic conductivity at $E = 0$ using the Nernst-Einstein equation based on the diffusion coefficient. The result shows a monotonically increasing trend, but the values are about two orders of magnitude smaller than those obtained from Eq.~\ref{sigmaE} (see Supplemental Materials [\onlinecite{supplement}]). This indicates that, rather than simple diffusion, cooperative motion among ions and spatial inhomogeneity within the crystal may play dominant roles in ionic conduction under electric field. 
Moreover, the ionic conductivity is found to be independent of the direction of the applied electric field. This is notable because, from the viewpoint of a Li$^+$ ion, the neighboring ions differ depending on crystallographic direction: in the (001) and (111) directions, neighboring ions are anions, whereas in the (101) direction, they are cations.
Since these directional differences should alter the local potential landscape, the observed isotropy of $\sigma$ suggests that macroscopic structural factors, rather than local crystallographic environments, dominate the conduction behavior.

\begin{figure}[htbp]
\centering
\includegraphics[width=16cm]{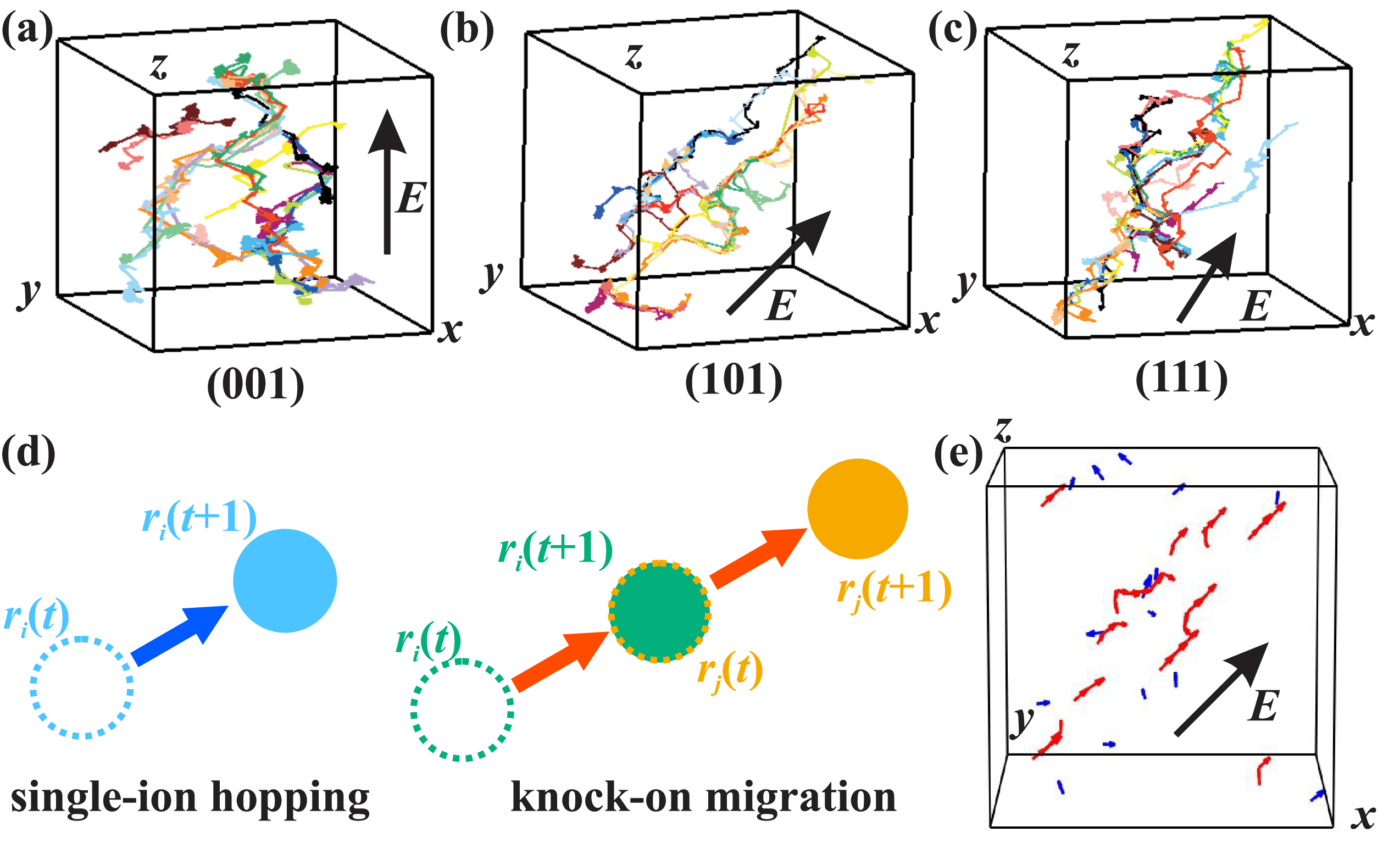}
\caption{
The trajectories of ions that moved more than a distance of $\Delta r = 3$ during a time interval $\Delta t = 300$, at $x = 1/3$ under an electric field $E = 2.1$ applied along the (a) (001), (b) (101), and  (c) (111) direction respectively.
Different colors represent different Li$^+$.
For clarity, only ions located near the center of the simulation box are shown.
Those mobile Li$^+$ follow the same trajectory aligned along the direction of the applied electric field.
(d) Schematic image of single-ion hopping and cooperative knock-on migration. 
(e) A snapshot of displacement vectors at $x = 1/3$ under an electric field $E = 2.1$ applied along the (101) direction during time interval $\Delta t = 1$. For clarity, displacement vectors crossing the simulation box boundaries are omitted. Red and blue arrows indicate cooperative knock-on migration and single-ion hopping, respectively.
}
\label{flow}
\end{figure}

To investigate the ion transport pathways within the crystal, we visualized the trajectories of ions exhibiting large displacements.
Figure~\ref{flow} (a)-(c) show only the trajectories of ions that moved more than a distance of $\Delta r = 3$ during a time interval $\Delta t = 300$, under an electric field $E = 2.1$ applied along the (001) (Fig.~\ref{flow}(a)), (101) (Fig.~\ref{flow}(b)), and (111) (Fig.~\ref{flow}(c)) direction at $x = 1/3$.
For clarity, only ions located near the center of the simulation box are displayed.
Each color represents a different ion, and all the visualized ions are Li$^+$.
The trajectories of Li$^+$ ions are generally aligned along the electric field direction and significantly overlap with each other, indicating the formation of specific conduction pathways.
Notably, the crystal structure remains intact, suggesting that Li$^+$ ions migrate cooperatively via a ‘knock-on’ mechanism, where the displacement of one Li$^+$ ion induces successive movement of neighboring ions without disrupting the lattice framework.
We also observe occasional transitions between distinct pathways, as noted in recent studies~\cite{Kihara2024, Weber2016, Yin2023}.
In future work, a quantitative analysis of how such inter-pathway hopping depends on the carrier concentration $x$ could provide deeper insights into the factors governing ionic conductivity in RSICs.
Here, to further elucidate the knock-on mechanism, we visualized the displacement vectors of Li$^+$. Figure~\ref{flow}(d) shows the schematic image of single-ion hopping and corporative knock-on migration and Figure~\ref{flow}(e) shows a snapshot of displacement vectors of Li$^+$ that moved more than $\Delta r = 0.7$ (approximately corresponding to a lattice hopping distance) within a time interval of $\Delta t = 1$, under an applied electric field of $E = 2.1$ along the (101) direction at $x = 1/3$ (see also the Supplementary Movie [\onlinecite{supplement}]). For clarity, displacement vectors crossing the simulation box boundaries are omitted. Red and blue arrows indicate cooperative knock-on migration and single-ion hopping, respectively.
We define a cooperative event between two ions $i$ and $j$ when $\min(|r_i(t+1)-r_j(t)|, |r_j(t+1) - r_i(t)|) \le \delta =0.34$ (ionic diameter of Li$^+$)~\cite{Donati1998, Morgan2021}. In such cases, the displacement vectors of both ions are shown as red arrows (see Ref.~\cite{Donati1998} for discussion on the choice of $\delta$).
Chains of 2-7 consecutive red arrows are typically observed within $\Delta t = 1$. By contrast, at $E = 0$, such cooperative knock-on migration events are rarely observed (see Supplementary Movie [\onlinecite{supplement}]). These results demonstrate that cooperative knock-on migration is induced only under electric field, which is consistent with the discrepancy between the $\sigma$ and $\sigma_{\mathrm{NE}}$. This suggests that ionic conduction in this system is governed not by single-ion hopping but essentially by cooperative knock-on migration.

\begin{figure*}[htbp]
\centering
\includegraphics[width=16cm]{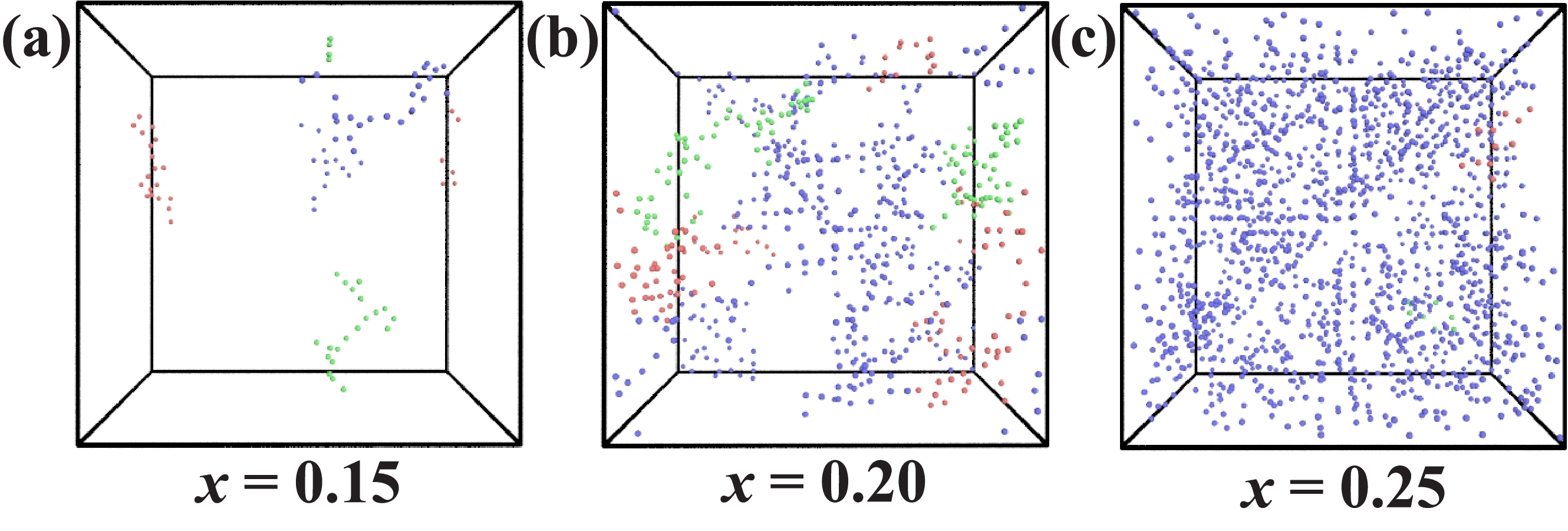}
\caption{Snapshots of the three largest Li$^+$ clusters at $x$ = (a) 0.15, (b) 0.2, and (c) 0.25.
Clusters are color-coded by size (blue, red, green).
Clusters are defined by Li$^+$ ions occupying nearest-neighbor cationic sites.
At $x = 0.15$, only small isolated clusters are observed.
At $x = 0.2$, clusters intermittently percolate, coinciding with the onset of the increase in $\sigma$.
At $x = 0.25$, a fully percolated single cluster spans the system.
}
\label{perco}
\end{figure*}

To examine the specific conduction pathways, we analyzed the spatial configuration of Li$^+$ ions.
Clusters were defined such that Li$^+$ ions occupying nearest-neighbor cationic sites were considered to belong to the same cluster.
Figure~\ref{perco} shows a visualization of the Li$^+$ clusters, where the three largest clusters are color-coded in blue, red, and green in descending order of size.
At $x = 0.15$, tiny clusters are formed (Fig.\ref{perco}(a)).
At $x = 0.2$, clusters occasionally percolate through the system, as shown in Fig.\ref{perco}(b), although they sometimes break apart, resulting in non-percolating configurations.
The second- and third-largest clusters are also relatively large at this composition, indicating that the system is near the onset of percolation.
This point coincides with the onset of the increase in $\sigma$, suggesting a strong correlation between $\sigma$ and Li$^+$ cluster percolation.
At $x = 0.25$, most of the Li$^+$ ions belong to a single large cluster that spans the entire system, indicating a fully percolated state (Fig.~\ref{perco}(c)).
Moreover, analysis of the temporal evolution of the clusters revealed that, on a microscopic scale, Li$^+$ continuously move in and out of the clusters, whereas on a macroscopic scale the percolating clusters remain largely unchanged and preserve their overall structure (see Supplementary Movie [\onlinecite{supplement}]).

Additionally, the diamond symbols (right axis) in Figure~\ref{sigma} show the $x$-dependence of the time-averaged length of the largest cluster, $L_c$.
Error bars represent the maximum and minimum values of $L_c$ over time.
$L_c$ is normalized by the maximum length of the simulation box, with $L_c = 1$ indicating that the cluster percolates throughout the system.
After $x = 0.25$, $L_c$ reaches 1 at almost all times, indicating a stable percolating network.
The large error bar observed at $x = 0.2$ reflects the intermittent breaking of the clusters.

Here, we compare the percolation theory for the FCC lattice, where cation sites in the rock salt crystal structure form an FCC lattice.
It is known that the threshold site-percolation probability for an FCC lattice is approximately 0.2~\cite{Stauffer1992}.
This value aligns well with the composition at which the onset of percolation is observed and $\sigma$ begins to increase.

\begin{figure}[htbp]
\centering
\includegraphics[width=8cm]{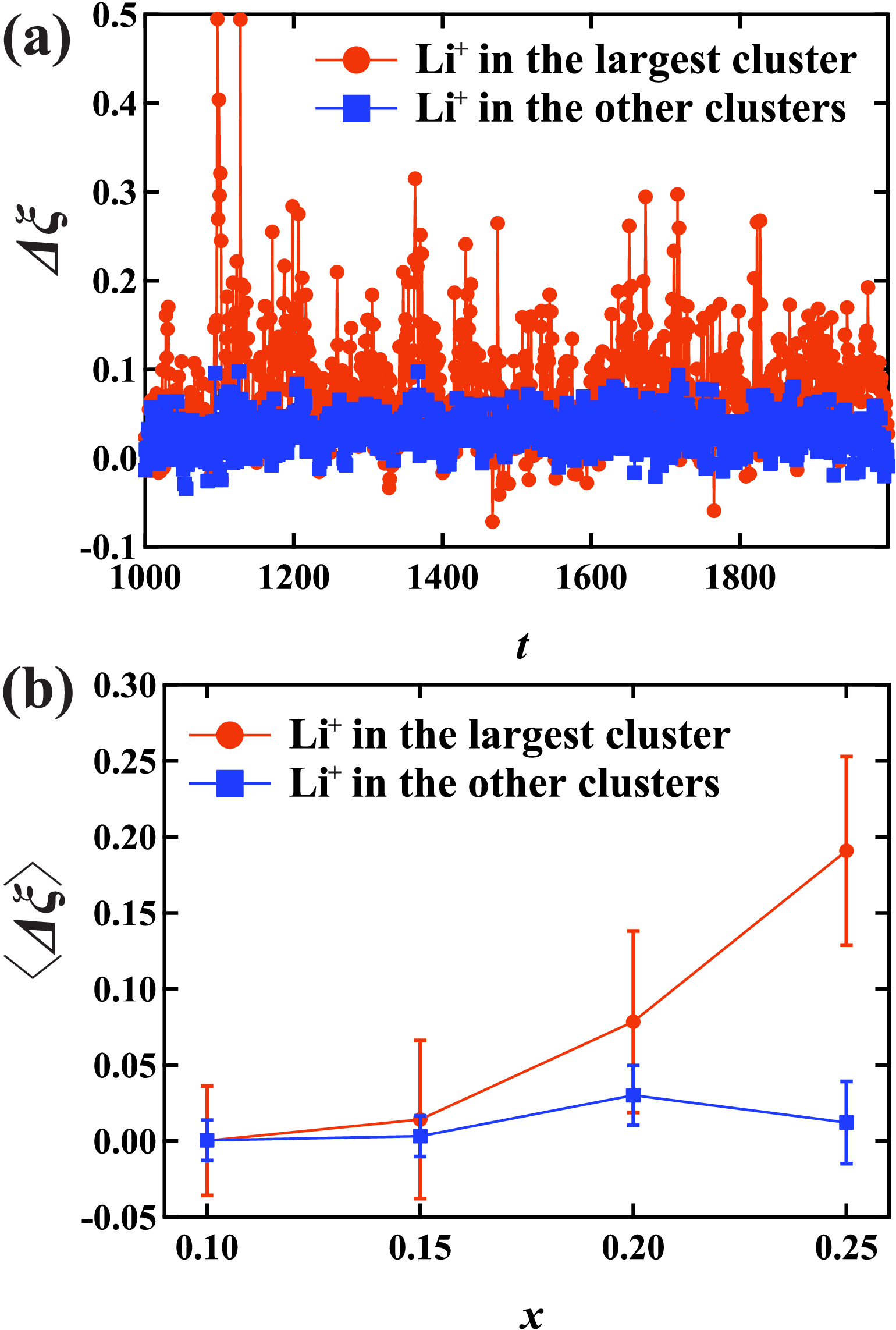}
\caption{
(a) Time evolution of the mean displacement $\Delta \xi$ of Li$^+$ ions along the $E$ direction over a time interval $\Delta t = 10$ at $x = 0.2$.
Circles and squares represent Li$^+$ ions belonging to the largest cluster and the other clusters, respectively.
At $x=0.2$, Li$^+$ ions in the largest cluster exhibit higher mobility, suggesting that the percolated clusters contribute to ionic conduction.
(b) $x$-dependence of the time-averaged displacement $\langle \Delta \xi \rangle$, with error bars indicating standard deviations.
Circles and squares correspond to Li$^+$ ions in the largest cluster and in the other clusters, respectively.
Above $x = 0.2$, $\langle \Delta \xi \rangle$ increases sharply for ions in the largest cluster, indicating that Li$^+$ percolation enhances ionic conduction.
}
\label{cluster}
\end{figure}

To confirm whether clusters contribute to ionic conduction, we compared the mobility of Li$^+$ ions belonging to the largest cluster with those in smaller clusters.
Figure~\ref{cluster}(a) shows the time evolution of the displacement $\Delta \xi$ of Li$^+$ ions, averaged over particles, along the $E$ direction over a time interval $\Delta t = 10$ at $x=0.2$.
Circles and squares represent the values of $\Delta \xi$ for Li$^+$ ions in the largest cluster and in the other clusters, respectively.
The $\Delta \xi$ is significantly larger for ions in the largest cluster, while ions in the smaller clusters exhibit negligible displacement.
Figure~\ref{cluster}(b) presents the $x$-dependence of the time-averaged displacement $\langle \Delta \xi \rangle$, with error bars indicating standard deviations.
Again, circles and squares correspond to Li$^+$ ions in the largest and smaller clusters, respectively.
For $x < 0.2$, $\langle \Delta \xi \rangle$ remains close to zero for both groups.
For $x \geq 0.2$ where the cluster percolates, $\langle \Delta \xi \rangle$ increases sharply for ions in the largest cluster, while remaining negligible for those in the other clusters.
These results indicate that ionic conduction is primarily facilitated by the percolated Li$^+$ clusters formed above the percolation threshold.

\begin{figure}[htbp]
\centering
\includegraphics[width=8cm]{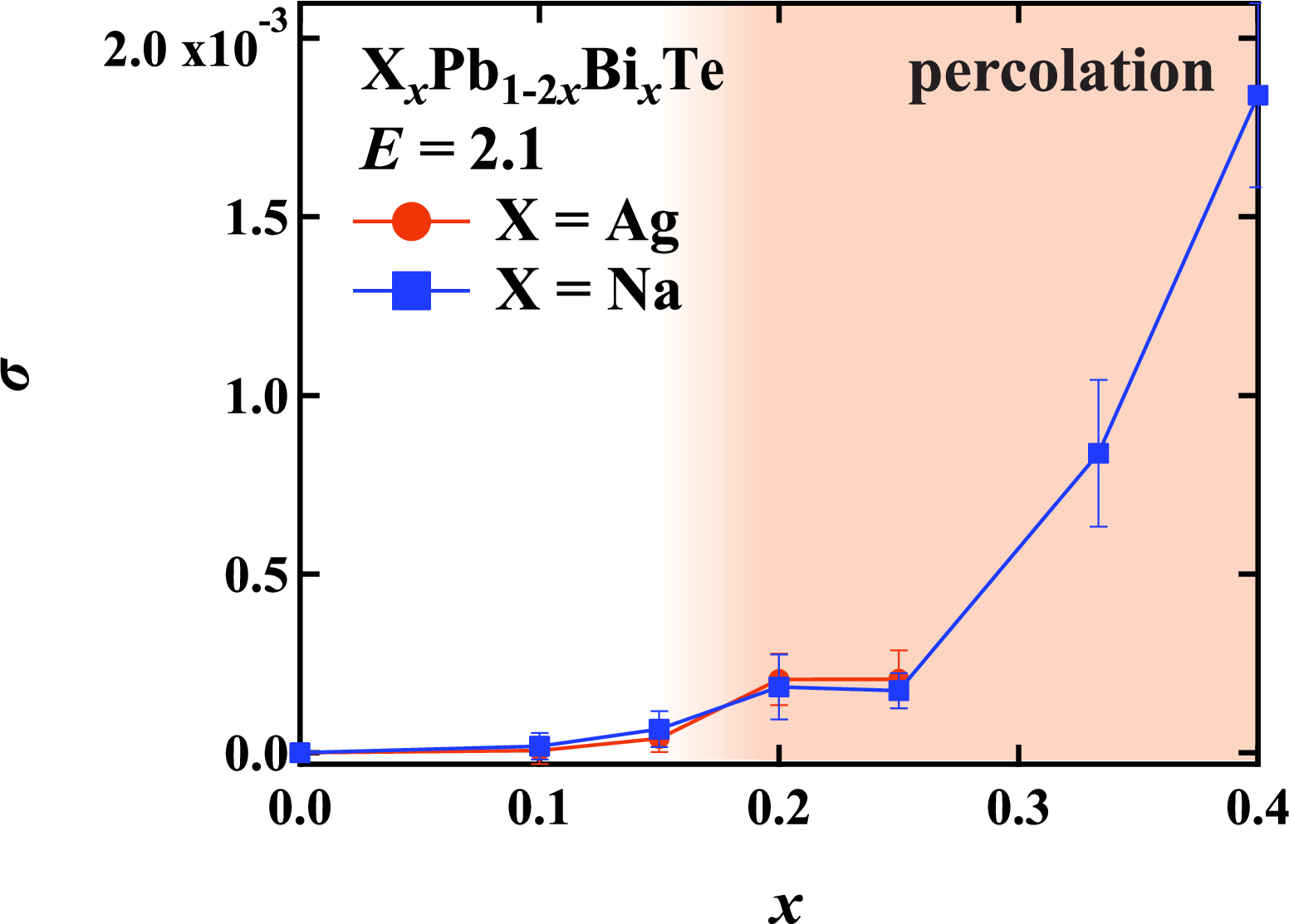}
\caption{The $x$-dependence of ionic conductivity $\sigma$ in X$_x$Pb$_{1-2x}$Bi$_x$Te (X = Ag, Na) under electric fields applied along the (101) direction. Circle and square correspond to X = Ag and Na, respectively. symbols and error bars indicate time-averaged values and their standard deviations. For Ag, crystals collapse when an electric field is applied for $x > 1/3$. 
Both roughly start to increase $\sigma$ at $x\sim0.2$. However, the absolute value of $\sigma$ is one order of magnitude smaller than in the Li$^+$ case, and the rise is more gradual.
}
\label{Na}
\end{figure}

Finally, we investigated whether ionic conduction occurs by a similar mechanism for cations other than Li$^+$. We performed additional simulations in which Li$^+$ was replaced by other monovalent cations (Na$^+$, K$^+$, and Ag$^+$) with ionic diameters of 0.461, 0.624, and 0.520, and masses of 0.180, 0.306, and 0.845, respectively. When an electric field was applied, the crystals including K$^+$ and Ag$^+$ were found to break down at $x > 0.2$ and $x > 1/3$, respectively, whereas the Na-substituted system retained its crystal structure. 
Figure~\ref{Na} shows that the $x$-dependence on ionic conductivity under an electric field $E = 2.1$ applied along the (101) in X$_x$Pb$_{1-2x}$Bi$_x$Te (X = Ag, Na). 
The ionic conductivity was observed to begin increasing at $x = 0.2$, as in the Li case, but its magnitude was about one order of magnitude smaller. These results suggest that the stability under electric field is reduced for cations other than Li, while the percolation threshold remains similar.

\section{Discussion}

Here, we discuss the importance of the anion-cation size difference in rock salt systems. In our system, the site percolation threshold coincides with the percolation of conduction pathways. However, when Li is replaced by a larger cation, the conductivity decreases, suggesting that a smaller size difference between the cation and anion makes ionic conduction less favorable.
In other rock salt systems, the percolation threshold does not directly coincide with site percolation. Instead, local structures~\cite{Lee2014,Yu2024} and chemical short-range order~\cite{Bin2020} strongly influence the energy barrier for Li hopping, and conduction pathways emerge preferentially along regions with low barriers. High conductivity arises when such low-barrier regions percolate through the lattice. Since the anions in those systems are mainly O$^{2-}$ (ionic diameter 2.80 $\mathrm{\AA}$), whereas our system contains Te$^{2-}$ (4.42 $\mathrm{\AA}$)~\cite{Shannon1976}, the difference in anion-cation size mismatch likely affects how conduction pathways form and why the effective percolation threshold differs. A more systematic investigation of this size-difference effect will be an important subject for future work.

Next, we examine why $\sigma$ exhibits no dependence on the $E$ direction.
The percolating Li$^+$ network forms uniformly throughout the crystal~\cite{Stauffer1992}, resulting in a macroscopic configuration without directional bias.
Due to the existence of isotropic Li$^+$ percolation pathways, when an external electric field is applied in any direction, ion conduction occurs efficiently along that field direction.
This finding highlights that, for device applications, controlling the crystallographic orientation is unnecessary to achieve high ionic conductivity.

We discuss the implications of our findings for materials design.
The ionic conductivity of Li$_{1/3}$Pb$_{1/3}$Bi$_{1/3}$Te at $T = 1.0$ (295 K) was calculated to be $6.8 \times 10^{-3}~\mathrm{S/cm}$ since unit ionic conductivity $\sigma_0 =  0.34~\mathrm{S/cm}$, comparable to that of liquid electrolytes~\cite{Nitta2015}.
While $\sigma$ increases with $x$, the melting point decreases (see Supplementary Information [\onlinecite{supplement}]), underscoring a trade-off between conductivity and thermal stability.
Our results demonstrate that ionic conduction is governed by percolation of the Li$^+$ network, providing a clear guideline: compositions exceeding the percolation threshold ($x \ge 0.2$) are promising candidates when balanced against stability requirements.
Given that Coulomb interactions dominate the transport mechanism, the concept should extend broadly to other ionic crystal systems such as high-entropy materials~\cite{Kasem2022}.
Future experimental verification will be important to further substantiate these results.

Finally, we note that when additional interactions play a significant role or when quantitative predictions of physical properties are required, more advanced approaches become necessary. In such cases, ab initio methods (including DFT) or machine-learning force fields would provide more realistic descriptions. We regard extending the present approach to these potentials as a natural and important next step toward bridging our simulations with material-specific applications.

\section{Summary}
In summary, this study investigates ionic conduction mechanisms in random substitutional crystals using molecular dynamics simulations of Li$_x$Pb$_{1-2x}$Bi$_x$Te.
We found that ionic conductivity ($\sigma$) sharply increases when the Li$^+$ concentration exceeds $x \approx 0.2$, corresponding to the onset of percolation of the Li$^+$ network within the crystal lattice.
Importantly, the percolating Li$^+$ clusters enable knock-on migration of ions without requiring Schottky defects, while preserving the crystal structure.
The ionic conduction is isotropic with respect to the direction of the applied electric field, owing to the random yet system-wide nature of the percolation network.
The achieved conductivity ($6.8 \times 10^{-3}~\mathrm{S/cm}$ at 295K) rivals that of liquid electrolytes, and our results offer a new design principle: by tuning the carrier ion concentration above the percolation threshold while maintaining thermal stability, highly conductive and chemically stable solid-state materials can be realized.
This percolation-driven conduction mechanism offers a physically grounded explanation for superionic behavior. Our results indicate carrier-ion percolation as a unifying and predictive design principle for next-generation solid electrolytes.

\section*{Acknowledgements}
R. I. was supported by MIYAKO-MIRAI Project of Tokyo Metropolitan University and JSPS Research Fellow Grant Number 24KJ1854. 
K. T. was supported by JSPS KAKENHI Grant Number 24K00594 and 25H01978. 
R. K. was supported by JSPS KAKENHI Grant Number 20H01874. 

\section*{AUTHORS CONTRIBUTIONS}
R.~K. conceived the project. R.~I. and K. T. performed the numerical simulations and analyzed the data. R.~I. and R.~K. wrote the manuscript.

\section*{COMPETING INTERESTS STATEMENT}
The authors declare that they have no competing interests. 

\section*{CORRESPONDENCE}
Correspondence and requests for materials should be addressed to R.~I. (ishikawa-rikuya@ed.tmu.ac.jp) and R.~K. (kurita@tmu.ac.jp).

\section*{Availability of Data and Materials}
All data generated or analyzed during this study are included in this published article and its supplementary information files.

\end{document}